# Inverse determination of effective mechanical properties of adhesive bondlines

Philipp Hass*, Falk K. Wittel, Miller Mendoza, Hans J. Herrmann, Peter Niemz

*Institute for Building Materials, ETH Zurich, Schafmattstrasse 6, 8093 Zurich, Switzerland*

*Corresponding author: +41 44 63 23250; +41 44 63 21174; phass@ethz.ch

## Abstract

A new approach for determining the effective mechanical bondline properties using a combined experimental-numerical modal analysis technique is proposed. After characterizing clear spruce wood boards, an adhesive layer is applied on the boards' surfaces. The shift of the eigenfrequencies resulting from the adhesive layer, together with information about the bondline geometry can then be used to inversely determine the mechanical properties of the adhesive layer using Finite Element Models. The calculated values for clear wood, as well as for the adhesive layer lie within reasonable ranges, thus demonstrating the method's potential.

## 1. Introduction

Bonding of wood elements is of paramount importance for the production of modern timber constructions. During the bonding process, adhesive penetrates into the porous wood structure along all accessible pathways and, depending on the adhesive properties, even into the micropores of the wood cell walls (Suchsland 1958; Kamke and Lee 2007; Konnerth et al. 2008; Hass et al. 2011). Simplified, a wood-adhesive bond can be split into five stacked layers consisting of pure adhesive in the center, an interphase of wood and cured adhesive as well as two wood parts on the outermost (Habenicht 2006). For clear wood, a vast amount of data is available (e.g. Kollmann and Côté 1968; Neuhaus 1981; Niemz 1993). The determination of the pure adhesive properties has been subject to several investigations as well (Clad 1964; Konnerth, Jager et al. 2006; Clauss et al. 2011). In contrast, the interphase properties are almost unknown. It can consist of entirely or partially adhesive filled tracheids, modified cell walls and a composite with adhesive, and cell-wall fragments from the production process. Nanoindentation was used to estimate the mechanical properties of penetrated cell walls and adhesive filled lumen (Konnerth and Gindl 2006). Unfortunately nanoindentation only measures very local material properties. For numerical models, however, it is necessary to have effective properties that consider the influence of the hardened adhesive on the wood properties, local defects due to curing stress such as cracks or voids (Hass et al. 2011), and geometric variations in a smeared way.

To identify the effective properties of the adhesive layer in the most mechanically meaningful way, a non-destructive approach seems appropriate. Eigenfrequency measurements have been used in the past to determine the properties of wood boards and to identify changes in their mechanical properties (Berner et al. 2007). In this work we want to further exploit the potential of using changes in the frequencies of certain eigenmodes to inversely identify the effective properties of the adhesive layer, using finite element simulations. Eigenmode shifts can be expected from the mass and effective properties of the adhesive layer as well as modifications of wood properties in the interphase zone. As a minimal model, the adhesive bond is simplified even further as a composition of only two layers with distinct differences in their material behavior: Solid wood with its strong orthotropic relationship between the three major directions, and a more or less isotropic layer consisting of adhesive and wood adhesive interface.

## 2. Material and Methods

To determine the mechanical properties of the wood-adhesive interphase, an inverse approach was chosen, in which experimental and numerical methods were combined. In a first step, the mechanical properties of clear wood were inversely determined. The procedure is repeated after an adhesive layer was applied on both sides of the board. This way the effect of the adhesive on the bending behavior and consequently also on the eigenfrequencies should be maximized. Warping of the samples, as it could be expected by a single-sided adhesive application should be avoided by the application of the adhesive on both sides and the consequent symmetrical property profile across the board thickness. In Figure 1a, the combined experimental and numeric procedure is schematized.

The spruce samples (*Picea Abies* (L.) Karst.) had the dimensions 320x125x5mm (orientation: longitudinal x radial x tangential). With this orientation, commonly referred to as standing growth rings, any effects of the growth ring curvature could be neglected, as the growth rings are aligned with the normal direction for bending. Additionally, this orientation is expected to give the highest frequency shifts during bending. Being a feasibility study, this investigation is limited to one adhesive system only. Urea-formaldehyde (UF) was chosen, since the typical massively cracked adhesive layer (see Figure 1b) should lead to effective properties that differ significantly from those of the pure adhesive. Only half of the amount of adhesive, recommended by the producers for one bondline (200g/$m^2$; solid content: 60%) was applied on each surface of the boards. To ensure an even distribution of adhesive on the boards, a uniform film was applied on a plastic sheet using an applicator with an adjustable slit (80µm). The sheets were then placed on the board to assure instantaneous contact and pressed at the recommended pressure of 0.7MPa. After curing, the samples were acclimatized at 20°C, 65% relative humidity (RH). To visually determine the bondline geometries, small samples were cut from the boards and prepared for analysis with an environmental scanning electron microscope (ESEM). This standard procedure included the embedding of the samples in epoxy, which allows for a better surface preparation (sanding and polishing) for the ESEM analysis. In Figure 1b a representative image is shown with the highlighted interface between clear wood and adhesive layer.

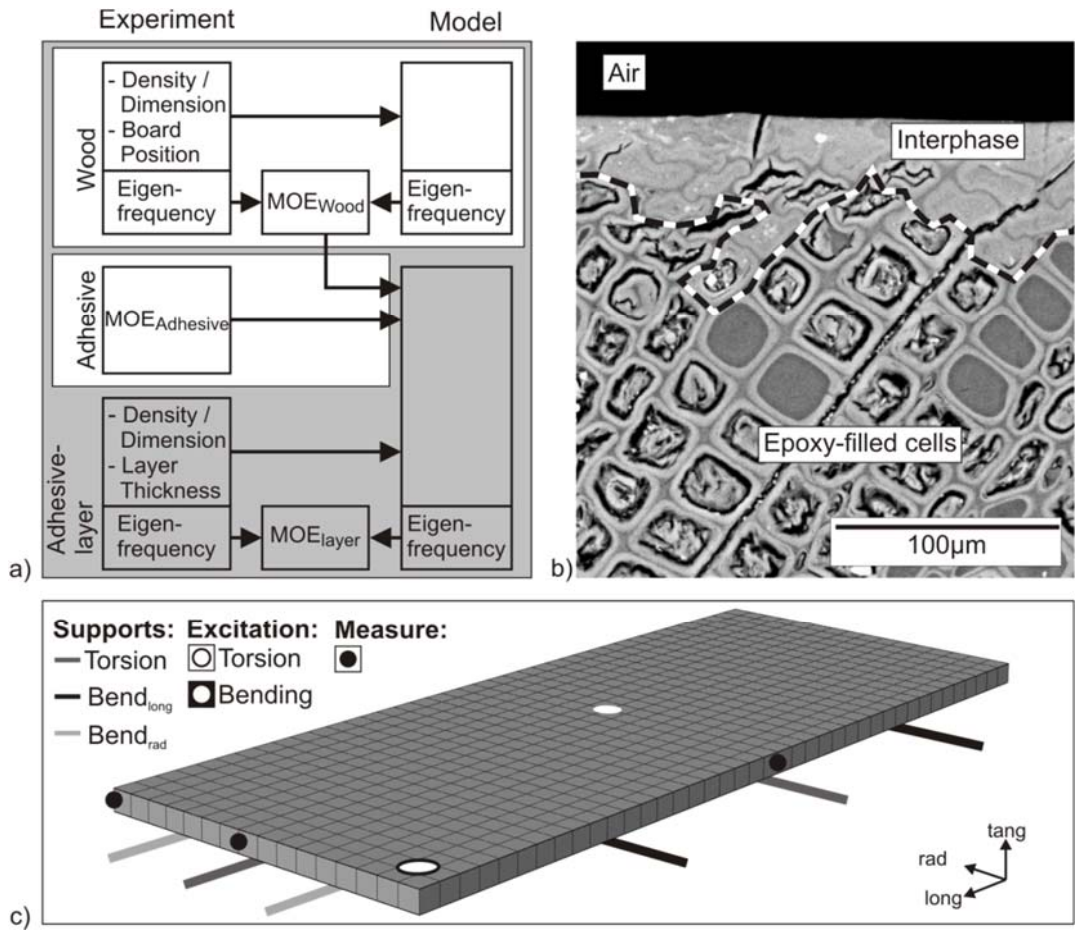

*Figure 1:* (a): Information flow between experiment and model. (b): Example for the determination of the UF-bondline geometry in spruce. (c): Experimental setup variants and mesh used in the simulations for the spruce wood samples with brick elements.

Literature values for the modulus of elasticity (MOE) of UF cover a wide range (Clad 1964: 1070-2590MPa; Bolton and Irle 1987: 3053-3743MPa). Therefore the used adhesive was mechanically characterized by compression tests on 5x5x5mm sized cubes that were cautiously cured and conditioned to avoid desiccation cracking. We measured a value of 3GPa for the UF adhesive used.

The frequency spectra of the samples were measured using a GRINDOSONIC impulse excitation measuring device MK5. For this purpose, the boards were placed on soft cellular plastic supports to minimize support-induced vibration of the boards. The impulse was given by a small ball-hammer at the points indicated in Figure 1c. Three different combinations of support setup and excitation location were chosen to enhance distinct vibrational modes of the boards: One bending oscillation each along the two small edges of the board and a torsion oscillation (Figure 1c). The spectra are transformed using a Fast-Fourier-Transformation (FFT) to extract the eigenfrequencies. After transformation, the signals of each measurement are normalized to the maximum peak in order to allow identification of the relevant modes.

The inverse determination of material properties consists of two parts: (i) The extraction of eigenvalues of the Finite Element system Abaqus and (ii) an optimization strategy to minimize the discrepancy between measured and extracted eigenfrequencies. From the linearized equation of motion of a discrete system with displacement vector $u$, namely

$$M\ddot{u} + Ku = p$$

for free vibration ($p=0$) and the harmonic ansatz for the displacements $u=\underline{u}e^{\omega t}$ leads to the Eigenvalue problem

$$(-\omega^2 M + K)\underline{u} = 0$$

with the global stiffness and mass matrix $K$ and $M$, the eigenvector $\underline{u}$ and the angular frequency of the undamped oscillation $\omega$. The non-trivial solution can be found through

$$\det\left|\omega^2 M + K\right| = 0.$$

However, direct determination of eigenvalues and respective eigenvectors of the K, M matrix pair is not possible. A number of iterative methods like the Lanczos eigensolver can extract eigenvalues from sparse symmetric systems (Argyris and Mlejnek 1991, Zienkiewicz et al. 2005). After extracting the eigenfrequencies $f_n$ that correspond to the n first bending and torsional modes, they are then compared to the experimental values $f_n^{exp}$ via an error function of the form

$$E_f(\%) = 100 \times \sqrt{\sum_n w_n \left(\frac{f_n - f_n^{exp}}{f_n^{exp}}\right)^2},$$

with $w_n$ as the weighting coefficient that is taken according to the accuracy of the experimental measurement for the respective frequency. This way it was assured that low frequencies that can be measured with high precision, are weighted higher ($w_n=0.2$) than higher eigenmodes ($w_n=0.1$), where the mapping of frequency to the respective mode is more difficult.

The underlying model has identical dimensions to the experiments and is discretized by volume elements with quadratic order (20 node brick element, C3D20R) but with reduced integration (Figure 1c) (Dassault Systems 2011). The plain wood is represented as a homogeneous region by an orthotropic material model, therefore by nine engineering constants $E_R$, $E_T$, $E_L$, $v_{RT}$, $v_{RL}$, $v_{TL}$, $G_{RT}$, $G_{RL}$, $G_{TL}$, and the density, all greater than 0. As a first step the sensitivity of all material parameters on the first eigenfrequencies is studied.

Starting values for MOE and shear modulus (G) were chosen to cover the entire range given in literature (Table 1). For the density, the measured values were taken as starting values (Table 1). Since the Poisson ratio variations merely change eigenfrequencies, they are fixed to values of $v_{RT} = 0.59$, $v_{RL} = 0.028$, and $v_{TL} = 0.02$ (values taken from Kollmann and Côté 1968; Neuhaus 1981), reducing the parameter space to seven dimensions. We made a variation of one property at a time in between reasonable boundaries given in Table 1 with equidistant increments in between.

The sensitivity study revealed that basically for variations of the MOE ($E_R$, $E_T$, $E_L$), a change of the bending frequency modes can be observed, while the torsion ones remain constant. The opposite is found for G ($G_{RL}$, $G_{RT}$, $G_{TL}$). Note that the important conditions for orthotropic materials, namely

$$|v_{LR}| < (E_L/E_R)^{1/2}, |v_{LT}| < (E_L/E_T)^{1/2}, |v_{RT}| < (E_R/E_T)^{1/2},$$

$$1 - v_{LR}v_{RL} - v_{RT}v_{TR} - v_{LT}v_{TL} - 2v_{RL}v_{TR}v_{LT} > 0$$

are always fulfilled. For each variant, the eigenfrequencies are calculated and the value of the error function is calculated using the experimentally determined frequencies. The value that produces the lowest error is chosen, and taken as the input value for the calculation of the next engineering constant. Thus, the values for the seven mechanical properties change in every completed cycle and, at the end of the ten cycles, the set of variables that optimize the error function is obtained. These values are introduced as initial parameters to the refinement stage. The refinement works in the same way, but instead of using upper and lower bounds with 10 steps in between, now all the values are allowed to vary +/-10% in 20 steps.

To calculate the eigenfrequencies of the adhesive covered boards, the model is extended by two layers with C3D20 elements of a uniform thickness δ. Their material is considered to be isotropic with fixed Poisson ratio (0.3) but variable MOE, density, and δ. The ranges for those variables were obtained from literature or measurements respectively (for ranges see Table 1). The thickness δ of the adhesive layer corresponds to the combination of pure adhesive and the wood-adhesive interphase. This layer was added to the pure wood samples without considering a reduced thickness of the wood samples by the interphase. Optimal values for the MOE, density and thickness are found using the identical procedure as described above.

## 3. Results and Discussion

From the huge number of eigenmodes, only six, namely the first two uniaxial bending and pure torsion modes, were selected (see Figure 2a), since those could be identified reliably, due to the experimental equipment used. The second bending mode along the radial direction (Bend$_{rad}$ 2) however, proved to be difficult to detect with the applied set-up. However when higher modes could occasionally be clearly identified, they were used for the optimization as well. The relative difference between the respective experimental and numerical eigenmodes is less than 2.25%.

In Table 1 the resulting material parameters for the clear wood boards are summarized. Note that they are in between known bounds from literature. As the boards' dimension in the tangential direction was much smaller than in the radial one, values for tangential direction cannot be found via inverse parameter identification, but are also insignificant for this work.

*Table 1: Summary of values for clear spruce wood, bulk UF adhesive and UF adhesive layer from literature, own measurements and calculations.*

| **Literature values** ||||||||||
|---|---|---|---|---|---|---|---|---|---|
| **Clear Wood** |||||||| **Bulk adhesive** | **Ref** |
| Density[a] [kg/m³] | MOE [MPa] ||| G [MPa] ||| | MOE [MPa] | |
| | L | R | T | LR | LT | RT | | | |
| 399 | 11700 – 13800 | 1680 – 1800 | 618 – 1170 | 617 – 642 | 587 – 615 | 51 – 53 | | | 1 |
| 470 | 10000 | 800 | 450 | 600 | 650 | 40 | | | 2 |

| | | | | | | | | |
|---|---|---|---|---|---|---|---|---|
| | 10163 – 12438 | 700 – 1069 | 386 – 710 | 610 – 856 | 590 – 787 | 51 – 97 | | 3 |
| | 11605 – 16716 | 687 – 903 | 392 – 638 | 618 – 746 | 500 – 853 | 29 – 39 | | 4 |
| 417 | 11933 | 803 | 412 | 733 | 612 | 41 | | 5 |
| | | | | | | | 1070 – 2590 | 6 |
| | | | | | | | 3053 – 3743 | 7 |

| **Measured values** | | | |
|---|---|---|---|
| **Clear wood** | **Bulk adhesive** | | **Adhesive layer**[c] |
| Density[a] [kg/m$^3$] | Density[a] [kg/m$^3$] | MOE [MPa] | Thickness (st. dev) [µm] |
| 461 | 1100 | 3000 | 48 (22) |
| 442 | | | 57 (15) |

| **Value-range for first calculation step** | | | | | | |
|---|---|---|---|---|---|---|
| **Clear wood** | | | | **Adhesive layer**[c] | | |
| Density[a,b] [kg/m$^3$] | MOE [MPa] | | | Density[a] [kg/m$^3$] | MOE [MPa] | Thickness [µm] |
| | L | R | T | | | |
| x*0.8 – x*1.2 | 7000 – 30000 | 300 – 2000 | 200 – 1500 | 500 – 3000 | 500 – 7000 | 10 –150 |
| | G [MPa] | | | | | |
| | LR | LT | RT | | | |
| | 360 – 1080 | 300 – 900 | 21 – 61 | | | |

| **Calculated Values (FEM)** | | | | | | | |
|---|---|---|---|---|---|---|---|
| **Clear wood** | | | | | **Adhesive layer**[c] | | |
| Density[a] [kg/m$^3$] | MOE [MPa] | | G [MPa] | | Density[a] [kg/m$^3$] | MOE [MPa] | Thickness [µm] |
| | L | R | LR | RT | | | |
| 478 | 13720 | 1164 | 911 | 45 | 1016 | 2567 | 71.5 |
| 440 | 14000 | 1326 | 601 | 41 | 792 | 2567 | 53.0 |

[a] Acclimatized at 20°C and 65% relative humidity;
[b]x stands for measured values of the respective sample;
[c]Adhesive layer consist of pure adhesive and wood-adhesive interphase;
[1]Keunecke et al. 2007; [2]Niemz 1993; [3]Bodig and Jayne 1982; [4]Kollmann and Côté 1968; [5]Neuhaus 1981; [6]Clad 1964; [7]Bolton and Irle 1987

In Figure 2b, the shift of the respective modes due to the adhesive is shown. It is striking that the bending mode in the radial direction changes most significantly. As the MOE of the boards in this direction is much smaller than in the longitudinal one, the adhesive layer has a higher influence on the MOE and therefore on the radial bending modes. An example of the frequency spectra with and without adhesive layer is given in Figure 2c.

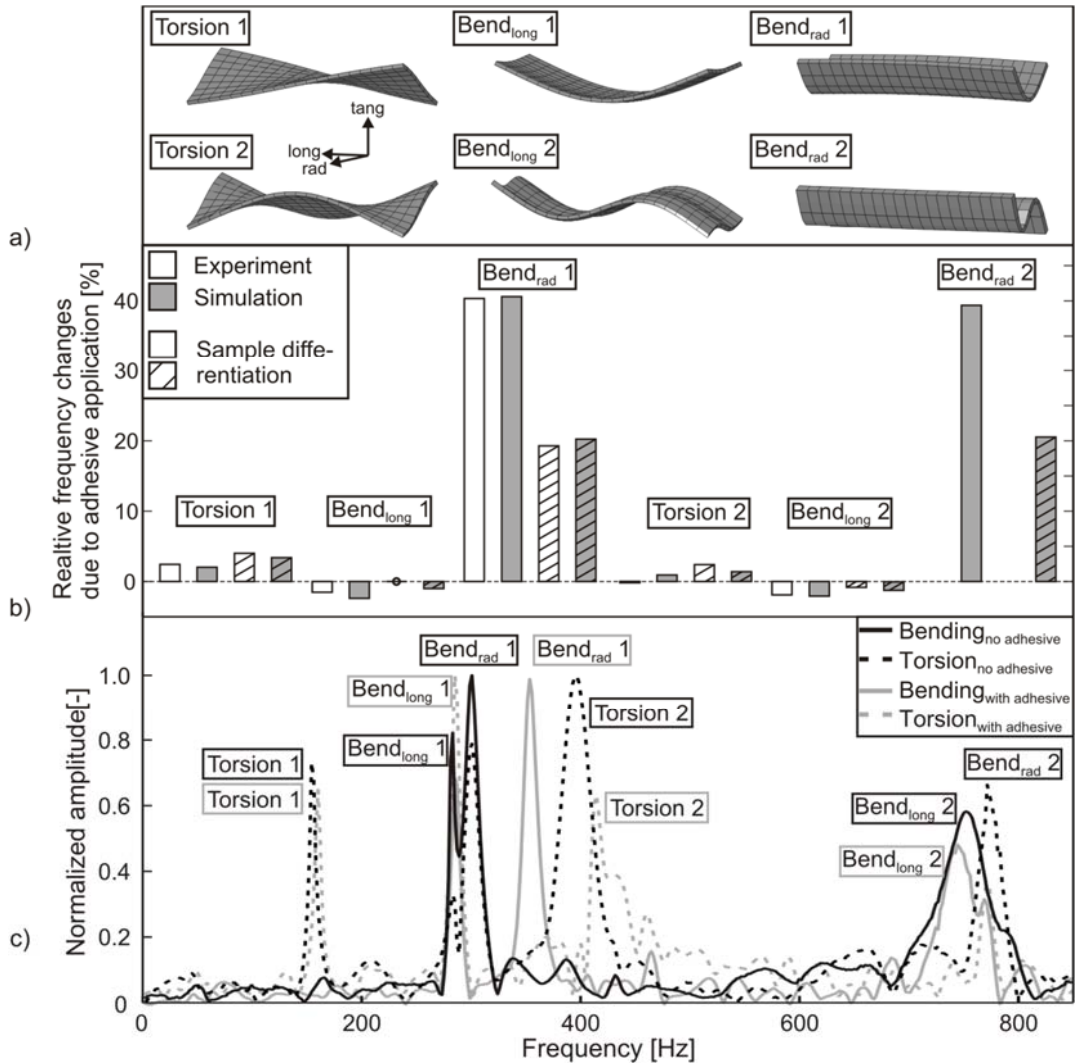

*Figure 2:* (a): Denotation of eigenmodes, which were used for comparison between clear wood samples and samples covered with adhesive layer. (b): Relative frequency shift in the single modes caused by the adhesive application. (c): Example of eigenmode-identification in clear spruce wood boards before and after application of UF adhesive layers. Shown are representative frequency bands, measured using different measurement setups, enhancing torsion and bending modes (see Figure 1c).

The resulting, inversely identified effective parameters of the adhesive layers are given in Table 1. It is interesting to find the effective MOE that lies in between the values of pure wood in the radial direction and the pure adhesive.

## Conclusions

With the proposed method it was possible to detect the eigenfrequency shifts caused by the application of an adhesive layer on clear spruce wood boards. Via inverse parameter identification, an estimate for the effective bond line properties can be made, that are needed for simulation purposes. All values are within reasonable ranges known from literature, showing the potential of the proposed approach. However several improvements could be made for future studies:

- With a more sophisticated setup, based for example on laser vibrometry, the eigenmode detection would be more accurate.

- With a quadratic board surface shape, overlapping modes should be avoided.

- To reduce the parameter space, mechanical tests would be helpful. This would be feasible by simple tension tests on strips produced from the boards. Bending tests would only be possible in the longitudinal wood direction since, as in the radial direction, the early wood section would falsify the results.

# Acknowledgements

Funding by the SNF- fund 200020_132662 Micromechanics of bondline failure and CRSI22_125184 Multiscale analysis of coupled mechanical and moisture behavior of wood as well as the provision of the adhesives by the companies Purbond/CH and Geistlich Ligamenta/CH is acknowledged.